\documentstyle[12pt,epsf,amsfonts,amssymb,amsbsy]{article}
\newcommand\VV{\setbox0=\hbox{V}\hbox{\rm V\raise\ht0
  \hbox to0pt{\hss\vbox to0pt{\hbox{v}\vss}}}}
\def\slashchar#1{\setbox0=\hbox{$#1$}           
   \dimen0=\wd0                                 
   \setbox1=\hbox{/} \dimen1=\wd1               
   \ifdim\dimen0>\dimen1                        
      \rlap{\hbox to \dimen0{\hfil/\hfil}}      
      #1                                        
   \else                                        
      \rlap{\hbox to \dimen1{\hfil$#1$\hfil}}   
      /                                         
   \fi}                                         %

\begin{document}

\vspace*{4cm}

\begin{center}
{\large \bf Two-loop anomalous dimensions for currents of
baryons with two heavy quarks in NRQCD.
}\\
\vspace*{5mm}
V.V. Kiselev, A.I. Onishchenko\\
{\sf State Research Center of Russia "Institute for High Energy Physics"} \\
{\it Protvino, Moscow region, 142284 Russia}\\
Fax: +7-095-2302337\\
E-mail: kiselev@mx.ihep.su
\end{center}
\begin{abstract}{
We present analytical results on the two-loop anomalous dimensions of currents
for baryons, containing two heavy quarks $J = [Q^{iT}C\Gamma\tau
Q^j]\Gamma^{'}q^k\varepsilon_{ijk}$ with arbitrary Dirac matrices $\Gamma$
and $\Gamma^{'}$ in the framework of NRQCD in the leading order over both
the relative velocity of heavy quarks and the inverse heavy quark mass. It is
shown, that in this approximation the anomalous dimensions do not depend on the
Dirac structure of the current under consideration.
}
\end{abstract}

\newpage
\section{Introduction}

The necessary feature of QCD applications to various fields of particle physics
is a study of a scale dependence for operators as it is governed by the
renormalization-group (RG). In the present paper we investigate the RG
properties of currents for baryons with two heavy quarks in the framework of
Non-Relativistic Quantum Chromodynamics (NRQCD) \cite{1},\cite{2} and its
dimensionally regularized version \cite{3}. In the two-loop approximation we
analytically calculate the anomalous dimensions of currents associated with the
ground-state baryons, containing two heavy quarks\footnote{We do not consider
the problems concerning the spectroscopy, decays and production mechanisms of
baryons with two heavy quarks. This can be found in \cite{A},\cite{B} and
\cite{C}, correspondingly.}. The dependence of QCD operators and matrix
elements on the relative velocity $v$ of heavy quarks inside the hadron as well
as on the inverse heavy quark mass $1/M_Q$ can be systematically treated in the
framework of justified effective expansions in QCD. So, we apply the expansion
in $1/M_Q$, as it was developed in Heavy Quark Effective Theory (HQET)
\cite{4,5,6} for operators, corresponding to the interaction of heavy quarks
with the light quark. For the heavy-heavy subsystem, the power tool is the
NRQCD-expansion in both the relative velocity and the inverse mass. Here we
consider the leading order in both $v$ and $1/M_Q$, that can serve as a good
approximation for the anomalous dimensions of currents under consideration.

The anomalous dimensions of composite operators can be desirably used in QCD
sum rules \cite{7}, which will allow us to evaluate the masses of these baryons
together with their residues in terms of basic non-perturbative QCD parameters.
For example, calculating the two-point correlators of baryonic currents in the
Operator Product Expansion (OPE) in NRQCD, we have to insert the anomalous
dimensions, obtained here in the static approximation, to relate the result to
QCD. 

This procedure is caused by a different ultraviolate behaviuor of loop
corrections in the full QCD and the effective theory. The latter contains the
divergences absent in QCD, since it was constructed in the way to provide
correct infrared properties of local QCD fields. The regularized quantities of
effective theory depend on the normalization point under the RG equations with
the corresponding anomalous dimensions. The ambiguity in the initial conditions
of such differential equations is eliminated by the matching to full QCD at a
scale, which is generally chosen as the heavy quark mass. The latter procedure
means, that using the effective theory, we can systematically take into account
the virtualities greater than the heavy quark mass. So, the knowledge of the
two-loop anomalous dimensions is also important, when one discusses the
matching of baryonic currents, obtained in this approximation with the
corresponding currents in full QCD. 

Our analysis in this paper is close to what was presented in \cite{9},
devoted to the baryons with a single heavy quark\footnote{We generally accept a
set of basic notations used in \cite{9}.}. While being very similar, these
analyses also have some differences, which we would like to
stress. The main technical obstacle of calculations is related to that
the kinetic term is thought to be a necessary ingredient in the quark
propagator for the evaluation of RG quantities in NRQCD, unlike to HQET,
\begin{equation}
\frac{1}{k_0+i\varepsilon}\longrightarrow \frac{1}{k_0-\frac{{\bf
k}^2}{2m}+i\varepsilon}.
\end{equation}
If a hard cut-off is used $(\mu\ll m)$, we can easily see that such
NRQCD-calculations can be performed just like in HQET, since $k^0\gg {\bf
k}^2/m$ in the ultraviolet regime. However, if the dimensional regularization
is used, the high energy modes $(k > m)$ are not explicitly suppressed and they
give non-vanishing contributions. This can be seen because the behavior of the
NRQCD propagator changes at energies greater than the mass. In spite of this,
one would like to use dimensional regularization because it keeps all of the
QCD symmetries and, moreover, the calculations are technically simpler. 

The difference between NRQCD and HQET can be explicitely highlighted in the
consideration of effective Lagrangian, derived to the tree level in the
$1/m$-expansion:
\begin{eqnarray}
\label{nrqcd}
{\cal L}_{\rm NRQCD} &=& \psi^\dagger \left(i D^0+\frac{\boldsymbol{D}^2}{2 m}
\right)\psi + \frac{1}{8 m^3}\,\psi^\dagger\boldsymbol{D}^4\psi-
\frac{g_s}{2 m}\,\psi^\dagger\boldsymbol{\sigma}\cdot \boldsymbol{B}\psi
\nonumber\\
&&\hspace*{-1.5cm}
-\,\frac{g_s}{8 m^2}\,\psi^\dagger\left(\boldsymbol{D}\cdot\boldsymbol{E}-
\boldsymbol{E}\cdot\boldsymbol{D}\right)\psi- 
\frac{i g_s}{8 m^2}\,\psi^\dagger\boldsymbol{\sigma}\cdot\left(
\boldsymbol{D}\times\boldsymbol{E}-\boldsymbol{E}\times\boldsymbol{D}\right)
\psi 
\nonumber\\[0.4cm]
&&\hspace*{-1.5cm}
+ O(1/m^3)+\,\,\mbox{antiquark terms}\,+ {\cal L}_{\rm light}
\end{eqnarray}
For a single heavy quark, interacting at low virtualities $D\sim
\Lambda_{QCD}$, the kinetic term is suppressed and can be treated
perturbatively. This results in the HQET prescription to the heavy quark
propagator. However, in the heavy-heavy system there is the Coulomb-like
interaction, wherein $D^0\sim \boldsymbol{D}^2/m\sim\alpha_s^2 m$. Therefore,
we must include the kinetic term into the initial "free" Lagrangian of NRQCD.
So, the loop corrections in $\alpha_s$ look to be different in HQET and NRQCD.
Nevertheless, the physical reason to distinguish these effective theories is
still the Coulomb-like corrections near the production threshold, which should
make no influence on the ultraviolate properties. We would note that the
question is, in a sense, analogous to that in the theory of massive gauge
fields in the spontaneously broken theories, where the explicite introduction
of mass seems to destroy good RG properties of massless vector fields (the
question was removed by the appropriate redefinitions of fields due to the
surviving the gauge invariance).

Several authors have addressed the similar problem of NRQCD in the connection
to matching calculations \cite{10}, and recently an appealing solution has been
proposed \cite{11}: it is claimed that the matching in NRQCD using the
dimensional regularization should be performed just like in HQET, namely the
kinetic term must be treated as a perturbation vertex:
\begin{equation}
\frac{1}{k_0-\frac{{\bf k}^2}{2m}+i\varepsilon} = \frac{1}{k_0}+\frac{{\bf
k}^2}{2m(k_0)^2}+...\label{2}
\end{equation}
The derivation is based on the appropriate redefinition of the heavy quark
field \cite{11}:
\begin{eqnarray}
Q&\longrightarrow &
[1-\frac{D_{\perp}^2}{8m^2}-\frac{g\sigma_{\alpha\beta}G^{\alpha\beta}}{16m^2}+
\frac{D_{\perp}^{\alpha}(iv\cdot
D)D_{\alpha\perp}}{16m^3}+\frac{gv_{\lambda}D_{\perp\alpha}G^{\alpha\lambda}}
{16m^3}\\
&& -i\frac{\sigma_{\alpha\beta}D_{\perp}^{\alpha}(iv\cdot D)
D_{\perp}^{\beta}}{16m^3}-i\frac{gv_{\lambda}\sigma_{\alpha\beta}D_{\perp}^{
\alpha}G^{\beta\lambda}}{16m^3}]Q,\nonumber
\end{eqnarray}
where the $\sigma$ matrices are projected by $P_v\sigma P_v$, 
$P_v = \frac{1+\hat v}{2}$ and $D_{\perp}^{\mu} = D^{\mu}-v^{\mu}v\cdot D$. 
The substitution converts the HQET Lagrangian to the NRQCD one, so that the
loop renormalization of perturbative terms is the same.

Here, we propose to use the same prescription for the heavy quark propagator as
it stands in (\ref{2}), not only in the matching procedure, but also for the
calculations of anomalous dimensions for the NRQCD currents in $\overline{\rm
MS}$-renormalization scheme. To support this point let us consider the matching
procedure in some details. The matching condition can be written down as
\begin{equation}
Z_{J,QCD}^{-1}Z_{2,QCD}^{on-shell}Z_{V,QCD}^{h.m.}\Gamma_{QCD}^{'} = C_0
Z_{2,NRQCD}^{on-shell}Z_{J,NRQCD}^{-1}\Gamma_{NRQCD}^{(0)},\label{3}
\end{equation}
\begin{equation}
Z_{V,QCD}^{h.m}\Gamma_{QCD}^{'} = \Gamma_{QCD}^{(0)},
\end{equation}
where $Z_V^{h.m.}$ denotes poles, associated with the hard momenta region for
the bare single-particle irreducible vertex $\Gamma_{QCD}^{(0)}$ in full QCD,
$Z_{J,QCD}$ and $Z_{J,NRQCD}$ are the renormalization constants of currents in
QCD and NRQCD, correspondingly, $Z_{2,QCD}$ and $Z_{2,NRQCD}$ include the
renormalization of wave functions, and, finally, $\Gamma_{NRQCD}^{(0)}$ denotes
the bare vertex in NRQCD. On this stage we use prescription (\ref{2}) for the
treating the heavy quark propagators. On the other hand, one can write the
following indentity
\begin{equation}
\Gamma_{QCD} = Z_{J,QCD}^{-1}Z_{2,QCD}^{\overline{\rm
MS}}Z_{V,QCD}^{h.m.}Z_{V,QCD}^{s.m.}\Gamma_{QCD}^{''},
\label{4}
\end{equation}
where we have collected all divergences in $Z$-factors and use the convention
of (\ref{2}) for the expansion of heavy quark propagators in powers of the
kinetic term. $Z_{V,QCD}^{s.m.}$ denotes the contribution from a small momenta
region. Calculating the contribution from the small momenta, we have to set
the external legs to be off-shell, in order to exclude the contribution from
the infrared region as it was done in the case of matching. To proceed further,
let us introduce the following definitions
\begin{eqnarray}
Z_{2,QCD}^{on-shell} &=& Z_{2,QCD}^{\overline{\rm MS}}Z_{inf.r.},\\
Z_{2,NRQCD}^{on-shell} &=& Z_{2,NRQCD}^{\overline{\rm MS}}Z_{inf.r.},
\end{eqnarray}
where $Z_{inf.r.}$ is a contribution to the wave-function renormalization from
the infrared region, which is the same in both theories. Using these notations
and the fact, that $Z_{2,NRQCD}^{on-shell} = 1$, we can rewrite Eq. (\ref{4})
as
\begin{equation}
\Gamma_{QCD} =
Z_{J,QCD}^{-1}Z_{2,QCD}^{on-shell}Z_{V,QCD}^{h.m.}Z_{V,QCD}^{s.m.}Z_{2,NRQCD}^{
\overline{\rm MS}}\Gamma_{QCD}^{''}.
\label{5}
\end{equation}
Now we can easily see from Eqs. (\ref{3}) and (\ref{5}), that the NRQCD
anomalous dimensions in the $\overline{\rm MS}$-renormalization scheme can be
computed in two ways: either from the matching condition (\ref{3}) or using the
HQET Feynman rules and setting the external legs off-shell in order to avoid
the infrared divergencies. We have explicitly checked this conjecture to
one-loop for the heavy-heavy vector current, however, for a full confidence we
feel a need for such a check in the two-loop approximation.

So, in our approach we exploit the same reasoning for the calculation of RG
quantities and work in the leading order of this expansion. Moreover, it is
theoretically sound, because in the $\overline{\rm MS}$-renormalization scheme
used, the anomalous dimensions of currents do not depend on the masses of
particles. The following fact also supports our claim: the values of Wilson
coefficients, calculated in the matching procedure, are directly connected to
the anomalous dimensions of operators multiplying these coefficients in the
Lagrangian.  And, finally, the high energy behavior in the effective theory
with several scales does not depend on relative weight of the lower scales.
Thus, we only need
\begin{equation}
m\gg |{\bf p}|, E, \Lambda_{QCD},
\end{equation}
where there is no matter what are relations between $|\bf p|$, $E$ and
$\Lambda_{QCD}$.

So, in our calculations we use the HQET propagators for the heavy quarks,
setting the quark momenta in a way to avoid infrared divergencies. As will be
explained in details below, the two-loop contribution to the anomalous
dimensions of currents under consideration consists of three parts. The first
corresponds to the set of graphs, wherein the two-loop contributions are
associated with one of the heavy-light subsystems. For this contribution we use
the result of \cite{9}. Then there is the subset of two-loop graphs that are
associated with the heavy-heavy system. The expression for this contribution is
a generalization of what was obtained in \cite{12}. And, finally, there are the
irreducible contributions, where the two-loops connect all three quark lines.
This contribution is calculated in this paper.  We evaluate the two loop
diagrams with the use of package, written by us on MATHEMATICA, and the
recurrence-relations in HQET \cite{13}.

This paper is organized as follows. In section 2 we discuss the choice of
currents for the baryons with two heavy quarks and give some comments on
the renormalization properties of composite operators under consideration.
In section 3 we furnish some remarks on the anomalous dimensions and present
the results on the one-loop anomalous dimensions. In section 4 we discuss
general features of two-loop renormalization procedure and present our two-loop
results.  We work in the $\overline{\rm MS}$-renormalization scheme throughout
the paper. As concerns the treatment of $\gamma_5$ we will show that the final
expression does not depend on the scheme used. Section 5 contains our
conclusion.

\section{Baryonic Currents}

The currents of baryons with two heavy quarks $\Xi_{cc}^{\diamond}$,
$\Xi_{bb}^{\diamond}$ and $\Xi^{\prime \diamond}_{bc}$, where $\diamond$ means
different charges depending on the light quark charge, are associated with the
spin-parity quantum numbers $j^P_d=1^+$ and $j^P_d=0^+$ for the heavy diquark
system with the symmetric and antisymmetric flavor structure, respectively.
Adding the light quark to the heavy quark system, one obtains
$j^P=\frac{1}{2}^+$ for the $\Xi^{\prime \diamond}_{bc}$ baryons and the pair
of degenerate states $j^P=\frac{1}{2}^+$ and $j^P=\frac{3}{2}^+$ for the
baryons $\Xi_{cc}^{\diamond}$, $\Xi_{bc}^{\diamond}$, $\Xi_{bb}^{\diamond}$ and
$\Xi_{cc}^{*\diamond}$, $\Xi_{bc}^{*\diamond}$,
$\Xi_{bb}^{*\diamond}$. The structure of baryon currents with two heavy quarks
is generally chosen as
\begin{equation}
J = [Q^{iT}C\Gamma\tau Q^j]\Gamma^{'}q^k\varepsilon_{ijk}.
\end{equation}
Here $T$ means transposition, $C$ is the charge conjugation matrix
with the properties $C\gamma_{\mu}^TC^{-1} = -\gamma_{\mu}$ and
$C\gamma_5^TC^{-1} = \gamma_5$, $i,j,k$ are colour indices and $\tau$ is a
matrix in the flavor space. The effective static field of the heavy quark is
denoted by $Q$. To obtain the corresponding NRQCD currents one has to perform
the above-mentioned redefinition of local field. But as we are working in the
leading order over both the relative velocity of heavy quarks and their inverse
masses, this local redefinition does not change
the structure of the currents.

Here, unlike the case of baryons with a single heavy quark, there is the only
independent current component $J$ for each of the ground state baryon currents.
They equal
\begin{eqnarray}
J_{\Xi^{\prime \diamond}_{QQ^{\prime}}} &=& [Q^{iT}C\tau\gamma_5
Q^{j\prime}]q^k\varepsilon_{ijk},\nonumber\\
J_{\Xi_{QQ}^{\diamond}} &=& [Q^{iT}C\tau\boldsymbol{\gamma}
Q^j]\cdot\boldsymbol{\gamma}\gamma_5
q^k\varepsilon_{ijk},\\
J_{\Xi_{QQ}^{*\diamond}} &=& [Q^{iT}C\tau\boldsymbol{\gamma}
Q^j]q^k\varepsilon_{ijk}+\frac{1}{3}\boldsymbol{\gamma}
[Q^{iT}C\boldsymbol{\gamma}
Q^j]\cdot\boldsymbol{\gamma} q^k\varepsilon_{ijk},\nonumber
\end{eqnarray}
where $J_{\Xi_{QQ}^{*\diamond}}$ satisfies the spin-3/2 condition
$\boldsymbol{\gamma}
J_{\Xi_{QQ}^{*\diamond}} = 0$. The flavor matrix $\tau$ is antisymmetric for
$\Xi^{\prime \diamond}_{bc}$ and symmetric for $\Xi_{QQ}^{\diamond}$ and
$\Xi_{QQ}^{*\diamond}$. The currents written down in Eq. (6) are taken in the
rest frame of hadrons. The corresponding expressions in a general frame moving
with a velocity four-vector $v^{\mu}$ can be obtained by the substitution of
$\boldsymbol{\gamma}\to \gamma_{\perp}^{\mu}=\gamma^{\mu}-\hat vv^{\mu}$.

Now we would like to give some comments on the renormalization properties of
these currents. As we have the only one light leg in this problem, all of
$\gamma$ matrices, which will appear in calculations, will stay on a single
side of our composite operators, not touching their Dirac structure. This will
lead to the fact that the anomalous dimensions of all our currents in this
approximation are the same, i.e. they do not depend on $\Gamma$-matrices in
(4). From this reasoning, we also can conclude that the result does not depend
on the $\gamma_5$ scheme used.

\section{Common notations in renormalization}

The local operators $O_0$ composed of bare physical fields contain the
ultra-violet divergences, which can be absorbed by the renormalization factors
$Z_O$, being a series in powers of coupling constant, so that $O=Z_O O_0$ is a
finite quantity, while the regularization parameters do not tend to peculiar
values. In the dimensional regularization using the $\overline{\rm MS}$-scheme
of subtractions in $D=4-2\epsilon$ dimensions \cite{tHVt}, $Z_O$ is expanded in
inverse powers of $\epsilon$, so that
\begin{equation}
Z=1+\sum_{m=1}^\infty\sum_{k=1}^m\left(\frac{\alpha_s}{4\pi}\right)^m
  \frac1{\epsilon^k}Z_{m,k}=1+\sum_{k=1}^\infty\frac1{\epsilon^k}Z_k.
\end{equation}
The dependence on the dimensionful subtraction point $\mu$ defines the
anomalous dimension of renormalized operator $O$
\begin{equation}
\gamma=\frac{d\ln Z(\alpha(\mu),a;\epsilon)}{d\ln (\mu)},
\end{equation}
where $a$ is the renormalized gauge parameter in the general covariant 
gauge (with a gluon propagator proportional to 
$-g_{\mu\nu}+(1-a)k_\mu k_\nu/k^2$) and $\alpha(\mu)$ is the renormalized 
coupling constant in four-dimensional space, so that
\begin{eqnarray}\label{conn}
\alpha_0=\alpha(\mu)\mu^{2\epsilon}Z_\alpha(\alpha(\mu),a;\epsilon),\qquad
a_0=aZ_3(\alpha(\mu),a;\epsilon),
\end{eqnarray}
and the corresponding $Z_{\{\alpha,3\}}$-factors determine the anomalous
dimensions, which are generally denoted by $\{-\beta,-\delta\}$, respectively.

The $\gamma$-quantities are finite at $D\to 4$, so we define the coefficients
of series
\begin{equation}\label{series}
\gamma=\sum_{m=1}^\infty\left(\frac{\alpha_s}{4\pi}\right)^m\gamma^{(m)}.
\end{equation}
One can check that \cite{PaTa}
\begin{equation}\label{anom}
\gamma=-2\frac{\partial Z_{1}}{\partial\ln\alpha_s},
\end{equation}
and for $k>0$
\begin{equation}\label{consistency}
-2\frac{\partial Z_{k+1}}{\partial\ln\alpha_s}
  =\left(\gamma-\beta\frac\partial{\partial\ln\alpha_s}
  -\delta\frac\partial{\partial\ln a}\right)Z_{k}.
\end{equation}
The latter provides the consistency condition, when the former produces a
simple extraction of the anomalous dimensions to the two-loop accuracy
\begin{equation}\label{andim12}
\gamma^{(1)}=-2Z_{1,1}\quad\mbox{and}\quad\gamma^{(2)}=-4Z_{2,1}.
\end{equation}

\subsection{One-loop result}

Consider the one-loop renormalization of currents of baryons with
two heavy quarks. In the $\overline{\rm MS}$-scheme with $D=4-2\epsilon$
space-dimensions we have the following squares of renormalization factors for
the bare quark fields:
\begin{equation}\label{ZqQ}
Z_q=1-a_0\frac{g_0^2C_F}{(4\pi)^2\epsilon},\qquad
Z_Q=1+(3-a_0)\frac{g_0^2C_F}{(4\pi)^2\epsilon},
\end{equation}
where we use the usual definitions for $SU(N)$, i.e.\ 
$C_F=(N_c^2-1)/2N_c$, $C_A=N_c$, $C_B=(N_c+1)/2N_c$, and $T_F=1/2$ for 
$N_c=3$, $N_F$ being the number of light quarks. 
One-loop $\overline{\rm MS}$-results for the factors $Z_\alpha$ and $Z_3$ 
have been given e.g.\ in \cite{PaTa}:
\begin{eqnarray}
Z_\alpha&=&1-\frac{\alpha_s}{4\pi\epsilon}
\left[\frac{11}3C_A-\frac43T_FN_F\right],\label{zal}\\ 
Z_3&=&1+\frac{\alpha_s}{4\pi\epsilon}
  \left[\frac{13-3a}6C_A-\frac43T_FN_F\right].\label{z3}
\end{eqnarray}
The bare current is 
renormalized by the factor $Z_J$:
\begin{equation}\label{renormj}
J_0=(Q_0^TC\Gamma\tau Q_0)\Gamma'q_0=Z_QZ^{1/2}_qZ_VJ=Z_JJ,
\end{equation}
which straightforwardly means that
\begin{equation}\label{gammaj}
\gamma_J=2\gamma_Q+\gamma_q+\gamma_V,
\end{equation}
i.e.  the anomalous dimension of the full 
current $J$ is a sum of three terms given by the renormalization 
of the light and heavy quark fields, and the renormalization of the vertex. 

For the vertex $(Q_0^TC\Gamma\tau Q_0)\Gamma'q_0$, we find
\begin{equation}\label{zgam1}
Z_V=1+\frac{\alpha_sC_B}{4\pi\epsilon}(3a-3),
\end{equation}
which results in
\begin{equation}\label{an1}
\gamma^{(1)}_V=-2C_B(3a-3).
\end{equation}
The one-loop anomalous dimensions $\gamma_q^{(1)}$ and $\gamma_Q^{(1)}$ are
equal to
\begin{equation}\label{gamma1legs}
\gamma_q^{(1)}=C_Fa,\qquad\gamma_Q^{(1)}=C_F(a-3).
\end{equation}
Thus, the one-loop anomalous dimension of the baryonic current is given by
\begin{equation}\label{an11}
\gamma_J=\frac{\alpha_s}{4\pi}\Big(-2C_B(3a-3)+3C_F(a-2)\Big)
  +O(\alpha_s^2).
\end{equation}

\section{Two-loop calculations}

In this section we apply the two-loop renormalization of the baryon current
with two heavy quarks in the $\overline{\rm MS}$-scheme and restrict ourselves
by the Feynman gauge. The two-loop anomalous dimensions of the quark fields are
given by \cite{13,Tara,Jone,Gime,JiMu}. 
\begin{equation}\label{gammaq}
\gamma^{(2)}_q=C_F\left(\frac{17}2C_A-2T_FN_F-\frac32C_F\right),\qquad
\gamma^{(2)}_Q=C_F\left(-\frac{38}3C_A+\frac{16}3T_FN_F\right).
\end{equation}

Since the baryonic currents are renormalized multiplicatively in the effective
theory\footnote{See discussion in ref.\cite{9}.}, the Dirac structure of vertex
repeats the Born-term. Technically we perform the calculations in terms of bare
coupling and gauge parameter, so that to isolate the two-loop contribution to
the anomalous dimension, we need also the one-loop result, wherein we have to
include the one-loop expressions written down through the renormalized
quantities $\alpha_s$ and $a$, which will add the contribution to the
corresponding $\alpha_s^2/\epsilon$-term. The procedure described leads to the
relations:
\begin{equation}\label{zvrel}
Z_{1,1}=V_{1,1},\qquad Z_{2,2}=V_{2,2},\qquad
  Z_{2,1}=V_{2,1}-V_{1,1}V_{1,0}.
\end{equation}
As expected, $Z_{2,1}$ has to include the one-loop contributions.

In the Introduction we have described three subgroups of two-loop diagrams for
the vertex, which can be expressed as
\begin{equation}
V_0=2V^{(hl)}_0+V^{(hh)}_0+V^{(ir)}_0,
\end{equation}
whose evaluation is presented in the rest of this section.

\subsection{The heavy-light subsystem.}
As for the problem of evaluation the bare proper vertex $V^{(hl)}$ of
composite operator $(qQ)$ with a massless quark field $q$ and the effective
static heavy quark field $Q$, we can easily see that the result does not depend
on the Dirac structure of the vertex. For this reason, in our calculations we
have used the result of \cite{9}, where this vertex was calculated to two-loop
order in the Feynman gauge $(a = 1)$ with the use of algorithm developed in
\cite{13}:
\begin{eqnarray}
V_{1,1}^{(hl)} = C_Ba,&&\quad V_{1,0}^{(hl)} = 0,\\
V_{2,2}^{(hl)} = C_B(\frac{1}{2}C_B - C_A),&&\quad V_{2,1}^{(hl)} =
-C_B(C_B(1-4\zeta (2))-C_A(1-\zeta (2))).\nonumber
\end{eqnarray}
Then from the relations~(\ref{zvrel}) one can calculate the coefficients
$Z_{n,k}$, which determine the two-loop anomalous dimension for the subset of
the heavy-light graphs 
\begin{equation}
\gamma_{(hl)}^{(2)} = C_B^2(4-16\zeta (2))-C_BC_A(4-4\zeta (2)).
\end{equation}
It is worth to note that this expression was calculated for antisymmetric
baryonic color configuration $q^iQ^jQ^k\epsilon_{ijk}$ unlike the case of
colour-singlet $\bar q^iQ^j\delta_j^i$ mesonic configuration. The expression
for the latter case can be obtained by substitution of $C_B\to C_F$, which
reconstructs the required results, as it was checkeds by authors of \cite{9}.

\subsection{The heavy-heavy subsystem.}

To evaluate this contribution we have used the results of \cite{12}, where the 
expression for the anomalous dimension of NRQCD mesonic vector
current was presented
\begin{eqnarray}
\gamma_J^M &=& 2\gamma_Q+\gamma_{(hh)} = \frac{d\ln
Z_J}{d\ln\mu}\\
&=& -C_F(2C_F+3C_A)\frac{\pi^2}{6}\Bigl(\frac{\alpha_s}{\pi}\Bigr)^2 +
O(\alpha_s^3).\nonumber
\end{eqnarray}

In our case we have the similar problem, but the different color
structure. Thus, following \cite{12} we can consider the matching of QCD vector
current with the antisymmetric color structure on the NRQCD one. Unlike the
meson case with the singlet-color structure, the QCD vector current with the
antisymmetric color structure need not be conserved, so we allow for its
renormalization. In terms of the on-shell matrix elements, the matching
equation can be written down as\footnote{Since the matching coefficient
contains only short-distance effects, the matching can be done by comparing the
matrix elements of these currents over a free quark-antiquark pair of the
on-shell quarks at a small relative velocity.} 
\begin{equation}
 Z_{2,QCD}Z_{J,QCD}^{-1}\Gamma_{QCD} =
C_0Z_{2,NRQCD}Z_{J,NRQCD}^{-1}\Gamma_{NRQCD} + O(v^2),
\end{equation}
where $Z_{J,QCD}$ has the following expression \cite{9}
\begin{eqnarray}
Z_{J,QCD} &=&
1-\frac{C_BC_F}{\epsilon^2}\biggl(\frac{\alpha_s}{4\pi}\biggr)^2+\frac{1}{
\epsilon}((C_B-C_F)\biggl(\frac{\alpha_s}{4\pi}\biggr)+\nonumber\\
&& (-\frac{1}{4}C_B(-17C_A+3C_B
+4(1+N_F)T_F)\\
&& +\frac{1}{4}C_F(-17C_A+3C_F+4(1+N_F)T_F))\biggl(\frac{\alpha_s}
{4\pi}\biggr)^2)\nonumber
\end{eqnarray}

The anomalous dimension of NRQCD current, obtained in this way, may be
used in the calculations of anomalous dimensions for the baryonic currents with
two heavy quarks, as it does not depend on the Dirac structure of the vertex.
The contributions of different two-loop diagrams with the antisymmetric
color structure of the vertex $q^iQ^jQ^k\epsilon_{ijk}$ in the notations of
\cite{12} are shown in Appendix. To obtain the anomalous dimension
$\gamma_{(hh)}^{(2)}$ of composite operator under consideration one has to
perform the following steps:

1) sum all of these contributions, including the one-loop term,
multiplyed by the two-loop QCD on-shell wave function renormalization
constant \cite{14}, $Z_{J,QCD}^{-1}$ and one-loop NRQCD-current renormalization
constant,

2) perform the one-loop renormalization of coupling and mass.

After these manipulations the coefficient at $\frac{1}{\epsilon}$ multiplied by
$-4$ will give us the sum $\gamma_{(hh)}^{(2)}+2\gamma_Q^{(2)}$. For the
two-loop anomalous dimension $\gamma_{(hh)}^{(2)}$ in the heavy-heavy
subsystem, we find the following result 
\begin{eqnarray}
\gamma_{(hh)}^{(2)} &=& -\frac{4}{3}C_B((-19+6\pi^2)C_A+4(\pi^2C_B+2N_FT_F)).
\end{eqnarray}

\subsection{The light-heavy-heavy irreducible vertex}

In this case one needs to calculate the three-quark irreducible vertex
$V_0^{(ir)}$. There are 8 diagrams in the two-loop order. We have shown four of
them in Fig. 1, the other four can be obtained by exchanging two heavy quark
legs. We set the heavy quarks off shell in order to avoid any infrared
singularities. Using the partial fractioning of the integrand in momentum
integrals and recurrence-relations of \cite{13}, we arrive to the following
expressions for the diagrams depicted on Fig.1 
\begin{eqnarray}
V_{0}^{(ir)[1]} &=& 2\cdot C_B^2\biggl(\frac{\alpha_s}{4\pi}\biggr)^2
\biggl(\frac{1}{2}\frac{1}{\epsilon^2} - (1+\frac{\pi^2}{3})\frac{1}{\epsilon}
+ \frac{216 + 35\pi^2 - 48\psi^{(2)}(1) - 96\psi^{(2)}(2)}{36}\biggr),\\
V_{0}^{(ir)[2]} &=& 0,\\
V_{0}^{(ir)[3]} &=& 2\cdot C_B^2\biggl(\frac{\alpha_s}{4\pi}\biggr)^2
\biggl(-\frac{1}{\epsilon^2} - \frac{2}{\epsilon} - 4 -
\frac{\pi^2}{6}\biggr),\\
V_{0}^{(ir)[4]} &=& 2\cdot C_B^2\biggl(\frac{\alpha_s}{4\pi}\biggr)^2
\biggl(-\frac{1}{\epsilon^2} + \frac{2}{\epsilon} - 4 -
\frac{3\pi^2}{2}\biggr),
\end{eqnarray}
where $\psi^{(n)}(z) = d^n\psi (z)/dz^n$, $\psi (z) = \Gamma^{'}(z)/\Gamma (z)$
and the factor of 2 accounts for the contributions of the remaining four
reflected diagrams not included in  Fig. 1. For the $Z$-factors and anomalous
dimension, we obtain
\begin{eqnarray}
Z_{22}^{(ir)} &=& -3\cdot C_B^2,\\
\gamma_{(ir)}^{(2)} &=& -4Z_{2,1}^{(ir)} = 8\cdot C_B^2(1+\frac{\pi^2}{3}).
\end{eqnarray}

\subsection{Anomalous dimension combined}

Now we are ready to calculate the anomalous dimension of baryonic currents with
two heavy quarks. As we have already said above it does not depend on the Dirac
structure of the current under consideration. Collecting the results for the
heavy-light, heavy-heavy and irreducible light-heavy-heavy vertices, we find
\begin{eqnarray}
\gamma_V^{(2)} &=& -\frac{4}{3}C_B((-13+30\zeta (2))C_A + 6(-2+6\zeta (2))C_B
+ 8N_FT_F).
\end{eqnarray}
And, finally, to obtain the full two-loop result for the anomalous dimension
one has to add the anomalous dimensions of heavy and light quarks. The result
is
\begin{eqnarray}
\gamma_J^{(2)} &=& \frac{1}{6}(-48(-2+6\zeta (2))C_B^2+C_A((104-240\zeta
(2))C_B-101C_F)\\
&& -64C_BN_FT_F+C_F(-9C_F+52N_FT_F)).\nonumber
\end{eqnarray} 
With this formula we are finishing our analytical calculations.

Landing to the SU(3) group of QCD, we get
\begin{eqnarray}
\gamma^{(1)} & = & -4 ,\\
\gamma^{(2)} & = & -\frac{254}{9}-\frac{152\pi^2}{9}+\frac{20}{9}N_F\approx
-194.909+2.222 N_F,
\end{eqnarray}
which indicate a rather strong sensitivity of those currents to the choice of
reference scale $\mu$.

\section{Conclusion}
We have calculated the two-loop anomalous dimensions of NRQCD baryonic currents
with two heavy quarks in the leading order in both the relative velocity of
heavy quarks and the inverse heavy quark mass. It is shown, that the results do
not depend on the Dirac structure of the currents and on the $\gamma_5$
prescription used in the calculations. These results will be useful for
derivation of QCD sum rules for baryons with two heavy quarks in the same
static approximation in both the leading and next-to-leading orders. We suppose
to address this problem in the nearby future.  

This work is in part supported by the Russian Foundation for Basic Research,
grants 96-02-18216 and 96-15-96575.

\section{Appendix}

In this appendix we present the generalization of expressions for the hard
contributions to the diagrams of Fig. 1 of \cite{12} with the antisymmetric
color structure of vertex evaluated at the threshold $q^2 = 4m^2$. Below you
can find the coefficients
of $(\alpha_s/\pi)^2(e^{\gamma_E}m_Q^2/(4\pi\mu^2))^{-2\epsilon}$

\begin{eqnarray}
D_1 &=& C_B^2
\bigg[~\frac{9}{32\epsilon^2}-(\frac{27}{64} +
\frac{5\pi^2}{24})\frac{1}{\epsilon}
- \frac{81}{128}-\frac{133\pi^2}{96} - \frac{5\pi^2\ln 2}{12}
-\frac{35\zeta(3)}{8}~\bigg],\\
D_2 &=& C_BC_F \bigg[~-\frac{3}{16\epsilon^2} - \frac{43}{32}\frac{1}{\epsilon}
+\frac{733}{192}+\frac{971\pi^2}{576}~\bigg],\\
D_3 &=& C_BC_A \bigg[~\frac{15}{32\epsilon^2} -
(\frac{5}{64}+\frac{\pi^2}{16})\frac{1}{\epsilon} +
\frac{715}{384}-\frac{319\pi^2}{576}-\frac{\pi^2\ln 2}{8}-
\frac{21\zeta(3)}{16}~\bigg],\\
D_4 &=& C_B(C_A-2C_B) \bigg[~(\frac{3}{16}-\frac{\pi^2}{16})\frac{1}{\epsilon}
-\frac{39}{32}-\frac{251\pi^2}{1152}-\frac{3\pi^2\ln 2}{8}-
\frac{31\zeta(3)}{16}~\bigg],\\
D_5 &=& C_B(C_A-2C_F)
\bigg[~-\frac{9}{32\epsilon^2}-\frac{19}{64}\frac{1}{\epsilon} +
\frac{761}{384}+\frac{1157\pi^2}{1152}+\frac{\pi^2\ln 2}{6}-
\frac{3\zeta(3)}{4}~\bigg],\\
D_6 &=& C_BT_FN_F
\bigg[~-\frac{1}{8\epsilon^2}+\frac{5}{48}\frac{1}{\epsilon}-\frac{355}{288}-
\frac{5
\pi^2}{48}~\bigg],\\
D_7 &=& C_BC_A \bigg[~\frac{19}{128\epsilon^2} -
\frac{53}{768}\frac{1}{\epsilon} +
\frac{6787}{4608}+\frac{95\pi^2}{768}~\bigg],\\
D_8 &=& C_BC_A \bigg[~\frac{1}{128\epsilon^2} + \frac{1}{768}\frac{1}{\epsilon}
+ \frac{361}{4608}+\frac{5\pi^2}{768}\bigg],\\
D_9 &=& C_BT_F \bigg[~-\frac{1}{4\epsilon^2} + \frac{13}{48}\frac{1}{\epsilon}
-\frac{145}{96}+\frac{5}{72}~\bigg].
\end{eqnarray}

\begin{center}
\begin{figure}[p]
\epsfxsize=16cm \epsfbox{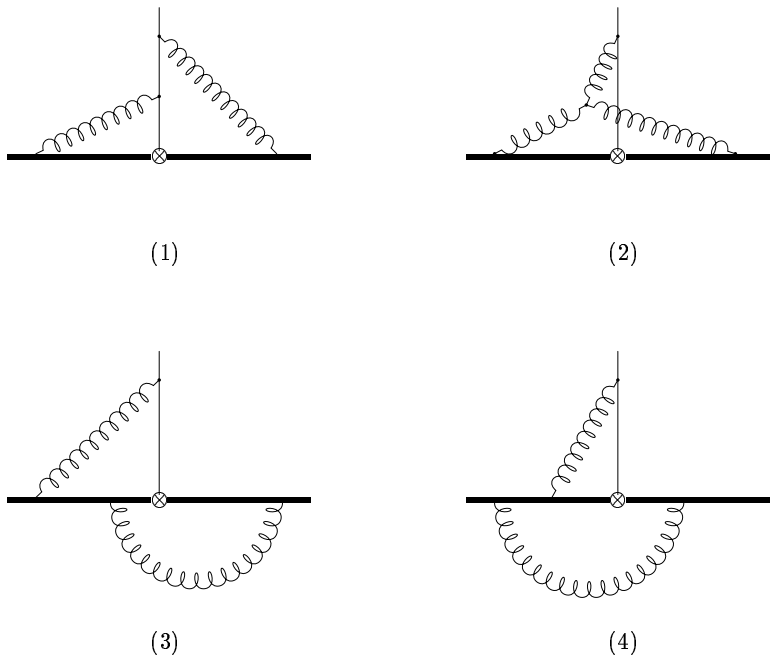}
\vspace*{-10cm}
\caption{The two-loop contribution to the light-heavy-heavy irreducible
vertex with the spinor lines directed outside.}
\label{Pic1}
\end{figure}
\end{center}
\end{document}